\documentclass[aps,pra,twocolumn,groupedaddress,superscriptaddress,bibnotes,amsfonts,longbibliography,citeautoscript,a4paper]{revtex4-2}
\usepackage{latexsym}
\usepackage{amsmath}
\usepackage{amssymb}
\usepackage{graphicx}
\usepackage[T1]{fontenc}
\usepackage[open]{bookmark}
\usepackage{hyperref}
\usepackage{orcidlink}
\hypersetup{colorlinks=true,allcolors=blue}
\usepackage{xcolor} 

\begin{document}

\title{
Quantum-informed plasmonics for strong coupling: the role of electron spill-out}

\author{Ida Juliane Bundgaard}
\affiliation{POLIMA---Center for Polariton-driven Light-Matter Interactions, University of Southern Denmark, Campusvej 55, DK-5230 Odense M, Denmark}

\author{Christian Nicolaisen Hansen}
\affiliation{POLIMA---Center for Polariton-driven Light-Matter Interactions, University of Southern Denmark, Campusvej 55, DK-5230 Odense M, Denmark}

\author{P.~Elli~Stamatopoulou\,\orcidlink{0000-0001-9121-911X}}
\affiliation{POLIMA---Center for Polariton-driven Light-Matter Interactions, University of Southern Denmark, Campusvej 55, DK-5230 Odense M, Denmark}
\author{Christos~Tserkezis\,\orcidlink{0000-0002-2075-9036}}
\email{ct@mci.sdu.dk}
\affiliation{POLIMA---Center for Polariton-driven Light-Matter Interactions, University of Southern Denmark, Campusvej 55, DK-5230 Odense M, Denmark}

\date{\today}

\begin{abstract}

The effect of nonlocality on the optical response of metals lies at the
forefront of research in nanoscale physics and, in particular, quantum
plasmonics. In alkali metals, nonlocality manifests predominantly as
electron density spill-out at the metal boundary, and as surface-enabled
Landau damping. For an accurate description of plasmonic modes, these
effects need be taken into account in the theoretical modelling of the
material. The resulting modal frequency shifts and broadening become
particularly relevant when dealing with the strong interaction between
plasmons and excitons, where hybrid modes emerge and the way they are
affected can reflect modifications of the coupling strength. Both
nonlocal phenomena can be incorporated in the classical local theory
by applying a surface-response formalism embodied by the Feibelman
parameters. Here, we implement surface-response corrections in Mie
theory to study the optical response of spherical plasmonic--excitonic
composites in core--shell configurations. We investigate sodium, a
jellium metal dominated by spill-out, for which it has been anticipated
that nonlocal corrections should lead to an observable change in the
coupling strength, appearing as a modification of the width of the mode
splitting. We show that, contrary to expectations, the influence of 
nonlocality on the anticrossing is minimal, thus validating the
accuracy of the local response approximation in strong-coupling
photonics.

\end{abstract}

\maketitle

\section{Introduction}\label{sec:introduction}

Plasmonics deals with the collective oscillations of free electrons in the
bulk or on the surface of metallic structures due to an external electromagnetic
(EM) excitation~\cite{PinesBohm,Ritchie,Economou1969,electronbeam_metalgrating}.
The quasiparticles of plasma oscillations on the surface of metallic
nanoparticles (NPs), called localized surface plasmons (LSPs), are known
for their ability to strongly enhance and confine the incident light in
subwavelength volumes~\cite{oulton_njp10}. Owing to these properties,
plasmonic NPs constitute ideal platforms for various applications, such as
in optical communications~\cite{Waele_2007}, photovoltaics~\cite{Pillai_2007}
and sensors~\cite{willets_arpc58}. To increase the practical benefit from
such systems, it has long been a trend to minimize the dimensions of
the NPs involved, in order to obtain a more dramatic confinement of
light. This goal is achievable as modern technologies allow the
fabrication and characterisation of metallic structures of ever
decreasing nanometric dimensions. At the extreme nanoscale of sizes
$\lesssim 10\ \mathrm{nm}$, the local response approximation (LRA),
based on the Drude model~\cite{jackson,Ashcroft_Harcourt1976}, or
on the experimentally measured permittivity of the bulk metal, can
no longer accurately predict the optical response of the system.

Quantum-informed plasmonics~\cite{drtechn} aims to incorporate nonlocal
effects of quantum-mechanical origin into the description of the
system~\cite{tame_2013}, including screening, Landau damping, and
electron spill-out~\cite{mortensen_nanoph10}. Screening, in the
case of a metallic NP, refers to the free electrons experiencing
a smaller restoring force, due to charge shielding from the bound electrons
in low-lying bands, leading to the electron density spilling inwards at the
assumed metal boundary~\cite{Hohenester}. Landau damping describes the
decay of the plasma oscillations into electron--hole pairs, and is enhanced
in small NPs due to collisions of conduction electrons with the metal
boundary~\cite{Li_2013,ElliArticle}. Electron spill-out describes the
fact that the electron density does not terminate abruptly, but rather
has a smoother profile, that spills out of the formal boundary of the
metal~\cite{MetalClusters}. One of the most widely used nonlocal models
that takes screening into account is the Hydrodynamic Drude model
(HDM)~\cite{ruppin_prl31,GBarton_1979,abajo_jpcc112}; by extending HDM
to the generalized nonlocal optical response (GNOR) model~\cite{GNOR},
we can additionally reproduce Landau damping. However, both HDM and GNOR
typically rely on a hard-wall boundary condition, which assumes that
the electron density is zero beyond the formal boundary of the
NP~\cite{melnyk_prb2,nonlocal_responses}, meaning that electron spill-out
is overlooked; self-consistent variants of HDM that can take this effect
into account have been proposed, but they remain computationally
heavy~\cite{HDM,yan_prb91,ciraci_prb93}.

Whether screening or spill-out is the dominant mechanism determining
the optical response of the metal depends on its work function. In alkali
metals, that are characterized by low work functions, spill-out is most
significant~\cite{SP_renormalization_effects,Performance_nonlocal,Qeffect_sodium}.
An elegant way to incorporate this effect in the description of the system
is offered by the Feibelman formalism, in which nonlocality is encompassed
by means of surface-response functions~\cite{feibd}. This framework
invokes new boundary conditions at the air--metal interface using the
surface-response parameters $d_\perp$ and $d_\parallel$, which
correspond to an induced surface charge and current density,
respectively~\cite{PlasmonParameters}. It thus allows the bulk response
of the plasmonic material to still be locally modelled, while introducing
a discontinuity due to the surface polarization~\cite{feibd}. In general,
the $d$-parameters are complex numbers with dimensions of length; the
real parts of the $d$-parameters express the position of the charge/current
centroid, whereas the imaginary parts account for surface-enabled Landau
damping~\cite{PRA}.

Due to the strong field enhancement and confinement that plasmonic NPs
achieve, they are often used as (open) nanocavities~\cite{Cavity_coupling},
in order to enhance the excitation and emission processes of nearby 
emitters~\cite{QD,Absorb_weak,Luminesscence_weak,Diode_weak,koenderink_ol35}.
When a cavity is coupled strongly to an emitter, modal hybridization
and reversible energy exchange can occur between the two
components~\cite{tserkezis_rpp83,Törmä_2015,Zhao_2022, Zengin_2015}.
One advantage of plasmonic nanocavities is the large control over
the plasmonic mode by the modulation of their size, which is enabled
by nanofabrication techniques~\cite{nanofab}. To achieve high coupling
strengths, plasmons are most commonly combined with excitonic
materials that sustain collective optical excitations, such as
low-dimensional semiconductors or organic dye
molecules~\cite{bellessa_prl93,chikkaraddy_nat535,stuhrenberg_nl18,geisler_acsphot6}.
By tuning the plasmonic mode and reducing the size of the metallic
NPs---and thus the mode volume, plasmon--exciton geometries show
strong potential for generating wide Rabi splitting, and thus
enabling the study of quantum effects at room
temperature~\cite{Hobson_2002,Bellessa_2009}.

The effect of screening and Landau damping in strongly-coupled, composite
core--shell NPs consisting of a noble metal and an excitonic material has
been studied earlier, within the HDM and GNOR models. In particular, it
was found that the width of the Rabi splitting is not affected by such
corrections. On the one hand, both hybrid modes shift in energy by the 
same amount; on the other hand, for such small NPs the increase of the
effective mode volume due to Landau damping is typically rather small,
and its influence on the anticrossing is counterbalanced by the nonlocal
blueshift of the modes~\cite{ACSphotonics}. Nevertheless, it was speculated
that the situation should be different in cases where spill-out is the
dominant quantum effect. In this work, the surface-response formalism
(SRF) framework is applied to determine whether electron spill-out indeed
affects the coupling strength in NPs consisting of sodium and excitonic
components. Specifically, we explore whether it influences the width of
the mode anticrossing, when compared to the LRA prediction. 

\section{Results and discussion}

In what follows, we study the optical response of core--shell NPs in
two different configurations, focusing on corrections pertaining to
electron spill-out and Landau damping. Sodium, in which spill-out is
non-negligible~\cite{Eckhardt1984,Townsend2012}, is the plasmonic material
considered throughout the paper, since its simple atomic structure has
rendered it the workhorse of theoretical quantum
plasmonics~\cite{zuloaga_nn4,zuloaga_nl9,teperik_prl110}. Within LRA,
sodium is modelled with the simple Drude model, expressing the relative
permittivity $\varepsilon$ as a function of angular frequency $\omega$
as
\begin{align}\label{Eq:Drude}
\varepsilon = 
1 - \frac{\omega_{\mathrm{p}}^{2}} 
{\omega \left(\omega + \mathrm{i} \gamma_{\mathrm{p}} \right)}
,
\end{align}
with plasma frequency $\omega_{\mathrm{p}}$ and damping rate
$\gamma_{\mathrm{p}}$, given as $\hbar \omega_{\mathrm{p}} =
5.89132$\,eV and $\hbar \gamma_{\mathrm{p}} = 0.1$\,eV. From now on,
all frequencies will appear normalized to $\omega_{\mathrm{p}}$.

\begin{figure*}[ht]
\centering
\includegraphics[width=\textwidth]{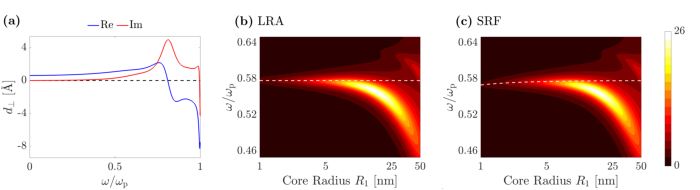}
\caption{(a) Real (blue line) and imaginary (red line) part of the
perpendicular Feibelman parameter $d_{\perp}$ for a flat sodium--air
interface~\cite{PlasmonParameters}. 
(b), (c) Contour maps of the extinction cross section (in logarithmic
scale), normalized to the geometric cross section $\pi R_{1}^{2}$, for
varying radius of a spherical sodium NP, obtained with
(b) LRA and (c) SRF, sharing a common color scale. The frequency of
the incident light $\omega$ is normalized to the plasma frequency
$\omega_{\mathrm{p}}$ of sodium in all panels. The white dashed line in
(b) marks the quasistatic frequency $\omega_{\mathrm{p}}/\sqrt{3}$;
in (c), the white dashed line gives the corresponding frequency
according to Eq.~(\ref{Eq:omega_Feibelman}).}
\label{fig:na_core}
\end{figure*}

The optical response of the NPs is investigated using Mie 
theory~\cite{mie_ap330}; the incident, scattered, and internal fields
are expanded in a sum of spherical waves of multipolar order $\ell$,
and the connection between their amplitudes is given by the scattering
Mie coefficients $T_{E\ell}$ and $T_{H\ell}$ (where $E, H$ stands for
transverse-electric and transverse-magnetic polarization, respectively).
The Mie coefficients are straightforward to calculate through the usual
pillbox boundary conditions at the particle (medium 1)--environment
(medium 2) interface, which, in the absence of surface charges and
currents, translate into continuity of the tangential components of the
electric and magnetic fields, $\mathbf{E}_{2 \parallel} - 
\mathbf{E}_{1 \parallel} = 0$ and $\mathbf{H}_{2 \parallel} - 
\mathbf{H}_{1 \parallel} = 0$~\cite{jackson}.

In SRF, the Feibelman $d$-parameters are introduced in a modified Mie
description. These parameters, typically calculated for an infinite flat
metal--dielectric interface, are defined as~\cite{feibd}
\begin{align}\label{eq:d_parameters}
d_{\perp} (\omega) =
\frac{\int \mathrm{d} x\; x \rho (x, \omega)}
{\int \mathrm{d} x\; \rho (x, \omega)}~,
\quad
d_{\parallel} (\omega) =
\frac{\int \mathrm{d} x\; x \partial_{x} J_{y} (x, \omega)}
{\int \mathrm{d} x\; \partial_{x} J_{y} (x, \omega)}
,
\end{align}
where $x$ is the direction normal to the interface, $\rho$ is the
induced charge density, and $\mathbf{J}$ the corresponding current
density. They can be incorporated into the boundary conditions of
the problem~\cite{PlasmonParameters},
\begin{subequations}\label{feib_bc}
\begin{align}
\mathbf{E}_{2 \parallel} -
\mathbf{E}_{1 \parallel} &= 
- d_{\perp} \nabla_{\parallel} 
\left( E_{2\perp} - E_{1\perp} \right)
,
\label{feib_bc_efield } \\
\mathbf{H}_{2 \parallel} -
\mathbf{H}_{1\parallel} &=
\mathrm{i}  \omega d_{\parallel} 
\left( \mathbf{D}_{2\parallel} -
\mathbf{D}_{1 \parallel} \right)
\times \mathbf{\hat{n}}
,
\label{feib_bc_hfield}
\end{align}
\end{subequations}
where $\mathbf{D}$ is the displacement field, $\nabla_{\parallel}$ is
the surface nabla operator, $\perp$ denotes the field component normal
to the interface, and $\mathbf{\hat{n}}$ the unit vector normal to the
interface. It should be noted that the use of Feibelman parameters for
curved interfaces imposes requirements on the dimensions of the system.
For a spherical particle of radius $R$, we require $R\gg \ell
d_{\perp,\parallel}$~\cite{PlasmonParameters}; the applicability of the
formalism can further be extended by considering nonlocal Feibelman 
parameter~\cite{babaze_oex30,babaze_nanoph12}. The modified Mie
coefficients used throughout this paper can be found in the Appendix. 

\begin{figure*}[ht]
\centering
\includegraphics[width=\linewidth]{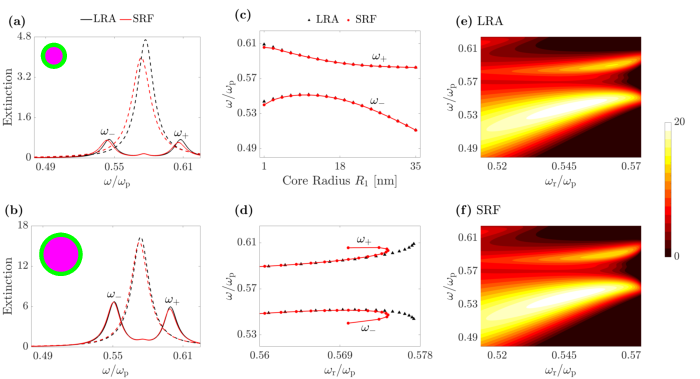}
\caption{(a, b) Extinction cross sections, normalized to the geometric
cross section $\pi R^{2}$, for a sodium core of radius (a) $2$\,nm  and
(b) $7.3$\,nm, both with an excitonic shell of $2$\,nm thickness obtained
from SRF (red solid line), compared to LRA (black solid line), showing
the two hybrid modes $\omega_{\pm}$. The dashed lines correspond to the
sodium core alone. 
(c, d) Hybrid mode resonance energies within SRF (red dots) and within LRA
(black triangles) as a function of (c) the sodium core radius and (d) the
sodium core resonance frequency $\omega_{\mathrm{r}}$, normalized to
$\omega_{\mathrm{p}}$. The red lines show the trend of the resonance
energies in SRF for both hybrid modes compared to LRA (black lines).
does not signify an increase in coupling strength but the shift of the
uncoupled resonance frequency $\omega_{\mathrm{r}}$.
(e, d) Logarithmically scaled extinction contour maps as a function of
resonance frequency $\omega_{\mathrm{r}}$ for (e) LRA and (f) SRF,
sharing a common color scale.}
\label{fig:na_core_exc_shell_fig}
\end{figure*}

\subsection{Surface Response Formalism for Sodium Nanoparticles}\label{sec:core}

To demonstrate the effect of spill-out on the optical response of plasmonic
NPs, we first consider a sodium sphere of radius $R_{1}$ embedded in air.
Following Ref.~\cite{PlasmonParameters}, the $d_{\perp}$ parameter is
given by Lorentzian fitting to time-dependent density-functional theory
calculations, where sodium was treated as a jellium of Wigner-Seitz radius
$r_{\mathrm{s}} = 4$~\cite{yan_2015}. A plot of the real and imaginary
part of $d_{\perp}$ is presented in Fig.~\ref{fig:na_core}a. We set
$d_{\parallel} = 0$, which is appropriate for charge-neutral
surfaces~\cite{feibd,PRA}.

Figs.~\ref{fig:na_core}b and c show the extinction cross sections of
a sodium NP surrounded by air, for varying radii $R_{1}$, calculated
within LRA and SRF, respectively. Up to a radius of around
$R_{1} = 25$\,nm, the only mode contributing to the extinction spectra
for both LRA and SRF is the electric dipolar (ED) LSP. As the radius
is decreased, the ED resonance energy in LRA converges to the quasistatic
limit $\omega_{1}^{\mathrm{(cl)}}/ \omega_{\mathrm{p}} = 1/\sqrt 3
\simeq 0.58$. This value is indicated with the dashed white line in
Fig.~\ref{fig:na_core}b, and captures the absence of retardation of the
free electrons in the NP, which experience the incident field as spatially
constant. In SRF, in which the electron density varies over the NP volume,
and thus the field that the NP experiences cannot be assumed constant, the
ED resonance energy deviates from the local quasistatic limit and follows
the relation 
\begin{align}\label{Eq:omega_Feibelman}
\omega_{1}^{\mathrm{(q)}} = 
\omega_{1}^{\mathrm{(cl)}}
\sqrt{1 - 2 d_{\perp} / R_{1}}
,
\end{align}
as indicated by the dashed white line in Fig.~\ref{fig:na_core}c. As the
NP dimensions increase, and retardation becomes important, higher multipole
contributions emerge. The electric quadrupolar (EQ) mode is seen in
Figs.~\ref{fig:na_core}b and c at approximately $0.61 \omega_{\mathrm{p}}$.
The effect of spill-out and Landau damping is most prominent for NP radii
below $5$\,nm, manifested in SRF as a red-shift and a broadening of the
plasmon resonance in the extinction spectrum, compared to the local model.

\subsection{Sodium Core with an Excitonic Shell}\label{sec: na_core_ex_shell}

Having now established the corrections to the plasmonic core alone, we
introduce an excitonic material to obtain strong coupling. Here, we
investigate a spherical sodium core of varying radius $R_{1}$, encapsulated
in an excitonic shell with a fixed thickness $W = 2$\,nm, embedded in air.
The excitonic shell is modelled with the Drude-Lorentz model,
\begin{align}\label{eq:DrudeLorentz}
\varepsilon_{\mathrm{exc}} (\omega) = 
1 - \frac{f \omega_{\mathrm{exc}}}
{\omega^{2} - \omega_{\mathrm{exc}}^{2} +
\mathrm{i} \omega\gamma_{\mathrm{exc}}}
,
\end{align}
with excitonic energy $\hbar\omega_{\mathrm{exc}} = 3.38$\,eV, damping rate
$\hbar \gamma_{\mathrm{exc}} = 0.06$\,eV, and oscillator strength $f = 0.02$~\cite{ExcitonParam}. The scattering Mie coefficients core--shell NP
are given by \eqref{eq:E_na_core_exc_shell} and \eqref{eq:H_na_core_exc_shell}
in the Appendix.

The spectra of the extinction cross section for NPs with $R_{1} = 2$\,nm and
$R_{1} = 7.3$\,nm sodium core are shown in Fig.~\ref{fig:na_core_exc_shell_fig}a
and Fig.~\ref{fig:na_core_exc_shell_fig}b, respectively. The SRF (LRA) extinction
spectrum is given by the red (black) line, with the dashed lines denoting the
extinction spectra for the sodium core in the absence of the excitonic shell.
All spectra are normalized to the geometric cross section $\pi R^{2}$, where
$R = R_{1} + W$ for the core--shell NP, and $R = R_{1}$ for the sodium core
alone. The interaction of the excitonic material with the LSP forms two hybrid
modes, as expected from literature~\cite{antosiewicz_acsp1}, with a mixed
plasmonic-excitonic chracter. Their energies are denoted as $\omega_{\pm}$
for the lower ($-$) and upper ($+$) hybrid mode, respectively. Between the
two hybrid modes there is a weak peak at $\omega/\omega_{\mathrm{p}} = 0.57$.
This is a geometric resonance of the charge oscillations that matches the
excitonic energy of the system, where the plasmonic core and excitonic shell
act as a uniform material~\cite{antosiewicz_acsp1}, and it does not participate
in the coupling.

For both core radii, the introduction of nonlocal effects through SRF produces
a redshift of both hybrid modes compared to the local model, albeit a more
significant shift for the upper hybrid mode than for the lower one. Additionally,
both modes in the nonlocal model experience broadening compared to the local
model, again more significant for the upper hybrid mode. The redshift and damping
of the hybrid modes is larger for the smaller core radius of $R_{1} = 2$\,nm in
Fig.~\ref{fig:na_core_exc_shell_fig}a, compared to $R_{1} = 7.3$\,nm in
Fig.~\ref{fig:na_core_exc_shell_fig}b. This can be explained by the electron
spill-out effectively increasing the core radius, allowing for LSPs of longer
wavelengths. This, in turn, introduces a detuning of the plasmon energy for
SRF compared to LRA, red-shifting the hybrid modes.

To better visualize the trends of the hybrid modes, a comparison of SRF
and LRA is provided in Figs.~\ref{fig:na_core_exc_shell_fig}c and d,
showing the resonance energies of the hybrid modes as functions of select
sodium-core radii and of select sodium-core resonance frequencies
$\omega_{\mathrm{r}}$, respectively. The resonance frequency
$\omega_{\mathrm{r}}$ is given by the LSP of the sodium core in the
absence of the excitonic shell for varying $R_{1}$. The radius of
the sodium core in Fig.~\ref{fig:na_core_exc_shell_fig}c varies from
$1$ to $35$\,nm, while the width of the excitonic shell is fixed at
$2$\,nm. We observe that for both hybrid modes, the differences in
energy between LRA and SRF hybrid modes are most significant for a
decreasing core radius $R_{1}$ smaller than about $R_{1} \approx 3$\,nm.
At a $1$\,nm core radius, which lies at the limits of validity of SRF,
the energy difference between LRA and SRF is $0.0035 \omega_{\mathrm{p}}$
for the upper mode, and $0.0037 \omega_{\mathrm{p}}$ for the lower
mode. For LRA, a decreasing radius corresponds to an increase in
$\omega_{\mathrm{r}}$. But for SRF, a decrease in core radius only
leads to an increase in $\omega_{\mathrm{r}}$ up to about
$0.574 \omega_{\mathrm{p}}$. At around this frequency, referencing
Fig.~\ref{fig:na_core}b, further decrease in $R_{1}$ now leads to a
decrease in $\omega_{\mathrm{r}}$. As both peaks of the SRF hybrid
modes are redshifted in energy compared to LRA, the energy difference,
and thus the coupling strength between the hybrid modes remains practically
unaffected in SRF compared to LRA at each core radius.

\begin{figure*}[ht]
\centering
\includegraphics[width=1\linewidth]{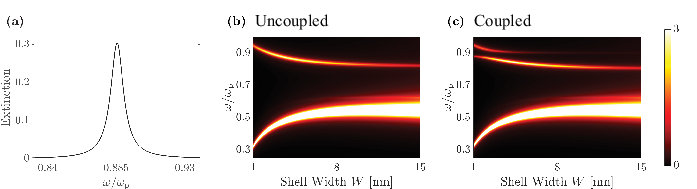}
\caption{(a) Extinction cross section of a spherical excitonic core
with radius $R_{1} = 5$\,nm embedded in air. 
(b, c) Extinction contour maps as a function of shell width for
(b) a vacuum core with a radius of $5$\,nm with a sodium shell, and 
(c) the combined excitonic core--sodium shell.
The color bar is shared for (b) and (c). All results are obtained within LRA.}
\label{fig:excitonic_core_plasmonic_shell_components}
\end{figure*}

When considering mode splitting, we commonly address the spectrum where
zero detuning between the two modes occurs, namely, where the resonance $\omega_{\mathrm{r}}$ of the sodium core matches the resonance of the
excitonic shell. Zero detuning in the local model occurs for a core radius
of $R_{1} = 7.3$\,nm, with the energy difference between the hybrid mode
peaks found to be $0.048 \omega_{\mathrm{p}}$. In the nonlocal model, zero
detuning occurs at a lower core radius than for LRA, at $R_{1} = 6.0$\,nm,
with the energy difference between hybrid modes being
$0.051 \omega_{\mathrm{p}}$ ($\simeq 300$\,meV). Comparing SRF to LRA, this
is a difference of $0.0027 \omega_{\mathrm{p}}$, with the larger anticrossing
occurring in the nonlocal model. With practically no change in the coupling
strength from LRA to SRF at each core radius, this change in anticrossing
is attributed to the fact that zero detuning occurs for two different radii.

Lastly, to examine the effect of resonance broadening due to Landau damping,
we consult the contour plots of Figs.~\ref{fig:na_core_exc_shell_fig}e and f,
where we plot the extinction cross section of the coupled system as a function
of the resonance energy $\omega_{\mathrm{r}}$, in LRA and SRF, respectively.
Clearly, no significant broadening can be observed, and the two plots are
nearly identical, since SRF reproduces a similar coupling strength. For high
frequencies of the incident light and low resonance frequencies
$\omega_{\mathrm{r}}$, we see also the EQ ($\ell=2$) resonance for the
sodium core, which, due to its large detuning to the excitonic energy,
does not participate in the coupling.

\subsection{Excitonic Core with a Sodium Shell}\label{sec: ex_core_na_shell}

We now investigate the inverted geometry, i.e. a spherical excitonic core
with a fixed radius $R_{1} = 5$\,nm, encapsulated in a sodium shell of
varying width $W$, and subsequently varying outer radius $R = R_{1} + W$.
The excitonic core is now modelled with an excitonic energy of $\hbar
\omega_{\mathrm{exc}} = 5.2$\,eV ($\omega_{\mathrm{exc}} = 0.88 \omega_{\mathrm{p}}$), while all other parameters remain the same as in section~\ref{sec: na_core_ex_shell}.
Because the new geometry has two plasmonic surfaces, Feibelman corrections must
apply on both. The nonlocal scattering Mie coefficients are now given by 
\eqref{eq:E_exc_core_na_shell} and \eqref{eq:H_exc_core_na_shell}, while for
the inner boundary we use \eqref{eq:E_na_core} and \eqref{eq:H_na_core} in
the Appendix. When considering the inner boundary of the sodium shell, it
should be noted that the spill-out occurs radially towards the centre of the
coordinate system. The Feibelman parameters, expressing lengths evaluated with
respect to the metal--environment interface, are directionally dependent, with
the positive direction previously assumed to be radially outwards. Therefore,
the sign of the Feibelman parameter shown in Fig.~\ref{fig:na_core} must be
inverted when dealing with the inner boundary of the sodium shell.

Fig.~\ref{fig:excitonic_core_plasmonic_shell_components}a shows the extinction
cross section of the excitonic core with a radius of $5\ \mathrm{nm}$, while
Fig.~\ref{fig:excitonic_core_plasmonic_shell_components}b shows a contour map
of the LRA extinction cross section of a sodium shell thickness $W$ ranging
between $1$ and $15$\,nm, in the absence of the excitonic core. The spectrum
of the excitonic core is a singular peak at $\omega_\mathrm{exc}/\omega_\mathrm{p}
\simeq 0.88$. The sodium shell has two distinct LSP modes; the higher energy mode
is a cavity mode, originating from LSPs at the inner surface of the shell,
while the lower mode is the NP mode, originating from the outer surface of the
shell~\cite{Tserkezis_2008}. A contour map of the LRA extinction cross section
for the sodium shell in the presence of the excitonic core is presented in 
Fig.~\ref{fig:excitonic_core_plasmonic_shell_components}c, showing an
anticrossing emerging between the cavity mode and the excitonic transition,
at the excitonic energy, indicating strong coupling. As the cavity mode is
the one participating in the coupling, in what follows we only consider this
mode when discussing the effect of the Feibelman corrections.

\begin{figure*}[ht]
\centering
\includegraphics[width=\linewidth]{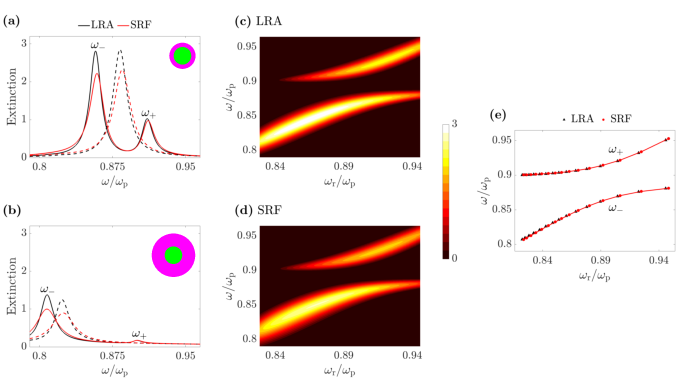}
\caption{(a, b) Extinction cross sections, normalized to the geometric cross
section $\pi R^2$, for a sodium shell of width (a) $3.14$\,nm and (b) $13.0$\,nm
around an excitonic core with a $5$\,nm radius. The hybrid modes $\omega_\pm$
for SRF (LRA) are shown as the solid red (black) lines. The dashed red (black)
lines are the extinction cross section of the sodium shell within SRF (LRA)
with an air core instead. 
(c, d) Logarithmically scaled extinction contour maps as a function of resonance
frequency $\omega_{\mathrm{r}}$ for (c) LRA and (d) SRF, sharing a common color
scale. 
(e) Hybrid mode resonance energies, shown as red dots (black triangles) within
SRF (LRA), as a function of the sodium shell resonance energy. The red lines
show the trend of the resonance frequencies of SRF for both hybrid modes.}
\label{fig:ex_core_na_shell_spectrum}
\end{figure*}

Figs.~\ref{fig:ex_core_na_shell_spectrum}a and b show the extinction
spectrum for the sodium shell with an air or excitonic core, for shell
widths $W = 3.14$\,nm and $W = 13.0$\,nm, respectively. When considering
nonlocal effects, the uncoupled cavity mode of the plasmonic shell
experiences damping and blueshift compared to the LRA spectrum, though
it can be seen that the blueshift is greater for the smaller shell width;
the upper hybrid modes on both graphs follow the same trend. Looking at
the lower hybrid modes in the two spectra, they both experience damping
in SRF compared to LRA, but in Fig.~\ref{fig:ex_core_na_shell_spectrum}b
we see a redshift for the lower hybrid mode. The opposite shifts are the
result of the nonlocality affecting both the coupling of the cavity mode
to the NP mode and the cavity mode alone. For this geometry, the electron
spill-out effectively increases the shell width compared to LRA, leading
to a spectral redshift of the cavity mode compared to LRA. At the same
time, at the inner shell boundary, the  electron spill-out effectively
decreases the inner radius $R_1$ of the shell, meaning that the LSPs in
the cavity for the SRF model will have shorter wavelengths than those
in the LRA model, leading to a blueshift of the SRF cavity mode spectrum
compared to LRA. In the quasistatic limit, we find the correction to the
electric dipole resonance frequency of the cavity mode to be
\begin{align}\label{Eq:omegacavity}
\omega_{1\mathrm{c}}^{(\mathrm{q})} =
\omega_{1\mathrm{c}}^{(\mathrm{cl})} 
\sqrt{1 - d_\perp/R_1}
,
\end{align}
where the quasistatic LRA cavity-mode resonance frequency is
$\omega_{1\mathrm{c}}^\mathrm{(cl)} / \omega_{\mathrm{p}} = 
\sqrt{2/3} \simeq 0.82$. These two nonlocal contributions result
in an overall blueshift of the lower mode at most shell widths,
such as the one shown in Fig.~\ref{fig:ex_core_na_shell_spectrum}a,
indicating that the contribution of the correction to the coupling
is more significant. However, at the larger shell widths, such as
the one shown in Fig.~\ref{fig:ex_core_na_shell_spectrum}b with an
overall redshift of the lower mode, the correction to the NP mode
becomes the more significant contribution.

To compare the mode splitting quantatively, we consider both models
at zero detuning, i.e. when the cavity mode of the sodium shell in the
absence of the excitonic core has resonance frequency $\omega_{\mathrm{r}}$
that equals the excitonic frequency $\omega_{\mathrm{exc}}$. The tuning of
the energy is done by varying the width of the sodium shell while keeping
the inner radius constant. Zero detuning occurs for the LRA shell, seen
in Fig.~\ref{fig:ex_core_na_shell_spectrum}a at width $W = 3.14$\,nm, where
the difference in energy between the resonance peaks is $312$\,meV. For SRF,
zero detuning occurs at a slightly larger width $W = 3.26$\,nm with a
difference in energy of $314$\,meV. Comparing the two values, it is
apparent that the energy difference between the corresponding coupling
strengths is negligible.

Focusing on the coupled cavity mode across the entire resonance frequency
spectrum, Figs.~\ref{fig:ex_core_na_shell_spectrum}c and d show the contour
maps of the extinction cross section for LRA and SRF, respectively, as a
function of the plasmonic resonance frequency $\omega_{\mathrm{r}}/
\omega_{\mathrm{p}}$ of the uncoupled cavity mode. The plasmonic resonance
energy is calculated as the resonance of the cavity mode of the sodium
shell in the absence of the excitonic core for a varying shell width $W$
and a fixed inner radius $R_{1} = 5$\,nm, and the shell width $W$ varies
from $1$ to $15$\,nm. On both panels, it can be seen that the splitting is
centred around the excitonic energy $\omega_{\mathrm{exc}}/\omega_{\mathrm{p}} 
\simeq 0.88$, as expected. Comparing Figs.~\ref{fig:ex_core_na_shell_spectrum}c
and d, the most notable difference is the broadening of the hybrid modes in
SRF compared to LRA due to Landau damping, especially for the lower mode.
Otherwise, the hybrid modes are still discernible in the strong coupling
regime.

To better visualize the splitting in the SRF and LRA models, Fig.~\ref{fig:ex_core_na_shell_spectrum}e shows the resonance energies of
the hybrid modes of select plasmonic resonance energies $\omega_{\mathrm{r}}$.
We see that, compared to the energy difference between the hybrid modes, the
prior described shifts of the SRF modes from the LRA modes become negligible.
This means that the energy split between the hybrid modes is practically not
affected by nonlocality for all the shown plasmonic resonance energies.

\section{Conclusion}
We examined the optical response of different core--shell configurations
of plasmonic--excitonic composites and compared the results obtained with
standard LRA to those produced with a nonlocal surface-response approach.
Specifically, we chose to work within the Feibelman framework, which is
ideal for describing the spill-out in alkali metals, together with
surface-enabled Landau damping. Even though energy shifts and resonance
broadenings of the hybrid modes are evident in the spectra, both models
generate very similar anticrossings in all configurations, since both
hybrid modes shift within SRF by the same amount. We thus found that
LRA provides a reliable estimation of the coupling strength, even when
the dimensions of metallic--excitonic systems are driven down to the
limits of nanofabrication capabilities.

\section*{Acknowledgments}
We acknowledge funding from VILLUM Fonden (Villum Investigator,
N. A. Mortensen, grant No.~16498).
The Center for Polariton-driven Light–Matter Interactions
(POLIMA) is sponsored by the Danish National Research
Foundation (Project No. DNRF165).

\section*{Appendix}
\label{Appendix}
Mie theory is used throughout this paper to calculate extinction
cross-sections and predict the optical response of the spherical NPs
presented. The coefficients and equations for the cross sections that
we have used are presented in this appendix. For all materials modelled,
the relative permeability is always assumed $\mu = 1$.

For a spherical NP with permittivity $\varepsilon_1$ and permeability $\mu_1$ embedded in a host medium with permittivity $\varepsilon_2$ and permeability $\mu_2$, the scattering Mie coefficients using an LRA framework are \cite{bohren_huffman, PrincipleOptics}
\begin{align}
    T_{E\ell}&=\frac{ j_\ell(x_{11})\Psi'_\ell(x_{21})\varepsilon_1- j_\ell (x_{21})\Psi '_\ell (x_{11})\varepsilon_2}{ h^{(1)}_\ell(x_{21})\Psi'_\ell(x_{11})\varepsilon_2- j_\ell (x_{11})\xi '_\ell (x_{21})\varepsilon_1}\label{eq:E_class}\\
    T_{H\ell}&=\frac{ j_\ell (x_{11})\Psi'(x_{21})\mu_1- j_{\ell}(x_{21})\Psi_\ell '(x_{11})\mu_2}{ h^{(1)}_\ell (x_{21})\Psi'(x_{11})\mu_2- j_{\ell}(x_{11})\xi_\ell '(x_{21})\mu_1},\label{eq:H_class}
\end{align}
where $x_{i1}=k_iR_1$ for wavenumbers $k_1$ and $k_2$ of the core and host material, respectively. The functions $\Psi_\ell'(z)$ and $\xi_\ell'(z)$ are the derivatives with respect to their argument $z$ of the Ricatti-Bessel functions $\Psi_\ell (z)=zj_\ell (z)$ and $\xi_\ell (z)=zh^{(1)}_\ell (z)$, where $j_\ell(z)$ and $h^{(1)}_\ell(z)=j_\ell(z)+\mathrm i y_\ell(z)$ are the spherical Bessel functions and spherical Hankel functions of the first kind, respectively (with $y_{\ell} (z)$ being the spherical Bessel function of the second kind, a.k.a Neumann function). 

We now introduce the Feibelman correction to a spherical NP of plasmonic material. The
scattering Mie coefficients in this SRF framework become~\cite{PlasmonParameters}
\begin{align}
&    T_{E\ell}=\frac{C^\mathrm{n}_{E\ell}+\bar{\varepsilon}[\bar d_\perp j_\ell (x_{21}) j_\ell (x_{11})+\bar d_\parallel \Psi_\ell ' (x_{21})\Psi_\ell ' (x_{11})]}{C^\mathrm{d}_{E\ell}-\bar{\varepsilon}[\bar d_\perp h^{(1)}_\ell (x_{21}) j_\ell (x_{11})+\bar d_\parallel \xi_\ell ' (x_{21})\Psi_\ell ' (x_{11})]}\label{eq:E_na_core}\\
&    T_{H\ell}=\frac{C^\mathrm{n}_{H\ell}+[\mu_2 x_{11}^2-\mu_1 x_{21}^2]\bar d_\parallel j_\ell(x_{21})j_\ell(x_{11})}{C^\mathrm{d}_{H\ell}-[\mu_2 x_{11}^2-\mu_1 x_{21}^2]\bar d_\parallel h^{(1)}_\ell(x_{21})j_\ell(x_{11})},\label{eq:H_na_core}
\end{align}
where $\bar{\varepsilon} = \varepsilon_1-\varepsilon_2$, $\bar d_\perp ={\ell(\ell +1)}d_\perp/{R_1}$ and $\bar d_\parallel = d_\parallel/{R_1}$. The coefficients $C^\mathrm{n}_{P\ell}$ and $C^\mathrm{d}_{P\ell}$ are the numerator and denominator from Eqs.~(\ref{eq:E_class}) and (\ref{eq:H_class}),
\begin{align*} 
    C^\mathrm{n}_{P\ell}=&j_\ell(x_{11})\Psi'_\ell(x_{21})p_1- j_\ell (x_{21})\Psi '_\ell (x_{11})p_2 \\
    C^\mathrm{d}_{P\ell}=&h^{(1)}_\ell(x_{21})\Psi'_\ell(x_{11})p_2- j_\ell (x_{11})\xi '_\ell (x_{21})p_1,
\end{align*}
where $p_m=\varepsilon_m$ if $P=E$ and $p_m=\mu_m$ if $P=H$. Note that if $d_{\perp,\parallel}=0$, the equations will become identical to Eqs.~(\ref{eq:E_class}) and (\ref{eq:H_class}).

We further add an excitonic shell to our plasmonic core NP; for a configuration with a non-plasmonic shell the scattering Mie-coefficients become
\begin{align}
    T_{E\ell }&=\frac{[\Psi_\ell'(x_{32})h^{(1)}_\ell (x_{22})\varepsilon_2-j_\ell (x_{32})\xi _\ell' (x_{22})\varepsilon_3 ]T_{E\ell }'+S^\mathrm{n}_{E\ell}}{[h^{(1)}_\ell(x_{32})\xi_\ell' (x_{22})\varepsilon_3-\xi_\ell' (x_{32})h _\ell (x_{22})\varepsilon_2 ]T_{E\ell }'+S^\mathrm{d}_{E\ell}}\label{eq:E_na_core_exc_shell}\\
    T_{H\ell}&=\frac{[\Psi_\ell'(x_{32})h^{(1)}_\ell (x_{22})\mu_2-j_\ell (x_{32})\xi _\ell' (x_{22})\mu_3 ]T_{H\ell }'+S^\mathrm{n}_{H\ell}}{[h^{(1)}_\ell(x_{32})\xi_\ell' (x_{22})\mu_3-\xi_\ell' (x_{32})h _\ell (x_{22})\mu_2 ]T_{H\ell }'+S^\mathrm{d}_{H\ell}},\label{eq:H_na_core_exc_shell}
\end{align}
where the arguments $x_{i2}=k_iR$, are in terms of the outer radius $R = R_1 + W$
and the wavenumbers $k_2$ and $k_3$ in the excitonic shell and host medium, air,
respectively. $T'_{P\ell}$ for polarization $P=E,H$ are the Mie coefficients for
the core, using \eqref{eq:E_class} and \eqref{eq:H_class} for LRA and
\eqref{eq:E_na_core} and \eqref{eq:H_na_core} for an SRF core. The coefficients 
$S^\mathrm{n}_{P\ell}$ and $S^\mathrm{d}_{P\ell}$ are given by
\begin{align*}
    S^\mathrm{n}_{P\ell}&=\Psi_\ell'(x_{32})j_\ell (x_{22})p_2-j_\ell(x_{32})\Psi_\ell' (x_{22})p_3\\
    S^\mathrm{d}_{P\ell}&=h^{(1)}_\ell(x_{32})\Psi_\ell' (x_{22})p_3-\xi_\ell'(x_{32})j_\ell (x_{22})p_2,
\end{align*}
where $p_m=\varepsilon_m$ if $P=E$ and $p_m=\mu_m$ if $P=H$. These coefficients are the LRA Mie coefficients evaluated at the shell-host boundary.

Inverting the geometry such that we have a non-plasmonic core and a plasmonic shell, the scattering Mie coefficients become
\begin{align}
    T_{E\ell }&=\frac{K^\mathrm{n}_{E\ell}+S^\mathrm{n}_{E\ell}+(\varepsilon_2-\varepsilon_3)(D^\mathrm{n}_{E\ell}+N^\mathrm{n}_{E\ell})}{K^\mathrm{d}_{E\ell}+S^\mathrm{d}_{E\ell}-(\varepsilon_2-\varepsilon_3)(D^\mathrm{d}_{E\ell}+N^\mathrm{d}_{E\ell})}\label{eq:E_exc_core_na_shell}\\
    T_{H\ell }&=\frac{K^\mathrm{n}_{H\ell}+S^\mathrm{n}_{H\ell}+(x_{22}^2\mu_3-x_{32}^2\mu_2)(D^\mathrm{n}_{H\ell}+N^\mathrm{n}_{H\ell})}{K^\mathrm{d}_{H\ell}+S^\mathrm{d}_{H\ell}-(x_{22}^2\mu_3-x_{32}^2\mu_2)(D^\mathrm{d}_{H\ell}+N^\mathrm{d}_{H\ell})},\label{eq:H_exc_core_na_shell}
\end{align}
where the coefficients $N^{n,d}_{P\ell}$ are 
\begin{align*}
    N^\mathrm{n}_{E_\ell}&=[\bar d_\perp h_\ell^{(1)} (x_{22})h_\ell^{(1)} (x_{32})+\bar d_\parallel \xi'_\ell (x_{22})\xi'_\ell (x_{32})]T_{E\ell}'\\
    N^\mathrm{d}_{E\ell}&=[\bar d_\perp h_\ell^{(1)} (x_{22})h_\ell^{(1)} (x_{32})+\bar d_\parallel \xi'_\ell (x_{22})\xi'_\ell (x_{32})]T_{E\ell}'\\
    N^\mathrm{n}_{H\ell}&=\bar d_\parallel h_\ell ^{(1)}(x_{22})j_\ell(x_{32})T_{H\ell}'\\
    N^\mathrm{d}_{H\ell}&=\bar d_\parallel h_\ell ^{(1)}(x_{22})h^{(1)}_\ell(x_{32})T_{H\ell}'
\end{align*}
We further introduce the coefficients $D^{n,d}_{P\ell}$ as
\begin{align*}
    D^\mathrm{n}_{E\ell}&=[\bar d_\perp j_\ell (x_{22})j_\ell (x_{32})+\bar d_\parallel \Psi'_\ell (x_{22})\Psi'_\ell (x_{32})]\\
    D^\mathrm{d}_{E\ell}&=[\bar d_\perp j_\ell (x_{22})h_\ell^{(1)} (x_{32})+\bar d_\parallel \Psi'_\ell (x_{22})\xi'_\ell (x_{32})]\\
    D^\mathrm{n}_{H\ell}&=\bar d_\parallel j_\ell(x_{22})j_\ell(x_{32})\\
    D^\mathrm{d}_{H\ell}&=\bar d_\parallel j_\ell(x_{22})h^{(1)}_\ell(x_{32}),
\end{align*}
which are terms also shown in \eqref{eq:E_na_core} and \eqref{eq:H_na_core}, they represent a similar correction to the shell--host boundary as previously considered for the core--host boundary. Lastly, we introduce the $K^{n,d}_{P\ell}$ coefficients
\begin{align*}
    K^\mathrm{n}_{P\ell}&=[\Psi_\ell'(x_{32})h_\ell^{(1)} (x_{22})p_2-j_\ell (x_{32})\xi _\ell' (x_{22})p_3 ]T_{P\ell }'\\
    K^\mathrm{d}_{P\ell}&=[h_\ell(x_{32})\xi_\ell' (x_{22})p_3-\xi_\ell' (x_{32})h _\ell^{(1)} (x_{22})p_2 ]T_{P\ell }',
\end{align*}
which are recognisable from \eqref{eq:E_na_core_exc_shell} and \eqref{eq:H_na_core_exc_shell} for the core--shell NP with non-plasmonic shell. Note that if $d_{\parallel,\perp}=0$ the equations will become identical to \eqref{eq:E_na_core_exc_shell} and \eqref{eq:H_na_core_exc_shell}.

The scattering $\sigma_\mathrm{sc}$, absorption  $\sigma_\mathrm{abs}$ and extinction $\sigma_\mathrm{ext}$ cross-sections for spherical NPs are calculated using Mie coefficients according to the following equations
\begin{align}
    \sigma_\mathrm{sc}&=\frac{2 }{(kR)^2}\sum_{\ell =1}^\infty (2\ell +1)(|T_{E\ell}|^2+|T_{H\ell}|^2)\\
    \sigma_\mathrm{ext}&=-\frac{2 }{(kR)^2}\sum_{\ell =1}^\infty (2\ell +1)\mathrm{Re}(T_{E\ell }+T_{H\ell })\\
    \sigma_\mathrm{abs}&=\sigma_\mathrm{ext} - \sigma_\mathrm{sc},
    \label{eq:cross_sections}
\end{align}
where $R$ is the outer radius, $k$ the wave vector of the host medium and $T_{E\ell}$ and $T_{H\ell}$ are the scattering Mie coefficients for the electric and magnetic multipoles, respectively.

\bibliography{references}

\end{document}